\documentclass[11pt]{article}
\pdfoutput=1
\usepackage{jheppub}

\preprint{\texttt{IFT-UAM/CSIC-25-40}}

\title{ \boldmath Cosmology inside a black hole: adding matter on the brane}

\author[a]{José L.F. Barbón,}
\author[a]{Ayan K. Patra,}
\author[a]{Juan F. Pedraza,}
\author[b]{Simon F. Ross}

\affiliation[a]{Instituto de Física Teórica UAM/CSIC, Calle Nicol\'as Cabrera 13-15, Madrid, 28049, Spain}
\affiliation[b]{Centre for Particle Theory, Department of Mathematical Sciences, Durham University, South Road, Durham DH1 3LE, U.K.}

\emailAdd{jose.barbon@csic.es}
\emailAdd{a.patra@csic.es}
\emailAdd{j.pedraza@csic.es}
\emailAdd{s.f.ross@durham.ac.uk}

\abstract{Braneworlds inside an AdS black hole provide simple models where a closed cosmology can be encoded in a dual field theory. Previous studies have focused on pure-tension branes, where the only matter in the cosmology is that dual to the holographic bulk. We consider models with additional explicit matter fields on the brane. Both for perfect fluid matter and for an axion field on the brane, we show that this can avoid a self-intersection problem in the Euclidean construction of such geometries in higher dimensions. }

\begin{document} 
\maketitle
\flushbottom

\section{Introduction}

\begin{figure}
    \centering
    \includegraphics[width=0.4\linewidth]{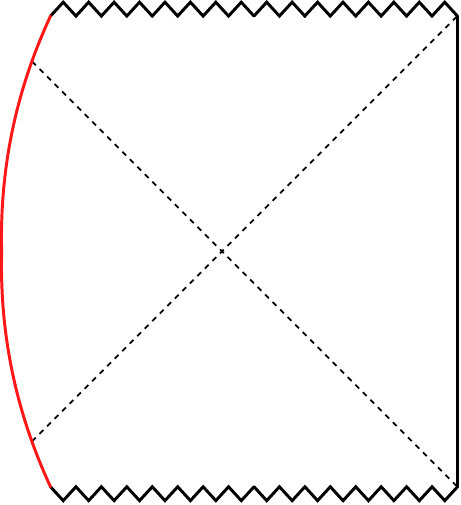}
    \caption{Penrose diagram of the scenario with an ETW brane behind the black hole horizon, providing an inner boundary of the spacetime.}
    \label{fig:penrose}
\end{figure}

Holography provides a very fruitful description of quantum gravity, but extending this to describe closed universes has presented significant challenges. One approach, initiated in \cite{Cooper:2018cmb} (see also \cite{VanRaamsdonk:2020tlr,VanRaamsdonk:2021qgv}), is to consider cosmology on a braneworld embedded in a higher-dimensional AdS bulk, with a dual description in terms of a field theory on the asymptotic boundary of the AdS space. In these scenarios we consider an asymptotically AdS$_{d+1}$ black hole, with a $d$-dimensional dynamical end of the world (ETW) brane behind the horizon, providing an inner boundary of the spacetime, as shown in figure \ref{fig:penrose}. The ETW brane emerges from the past singularity and terminates at the future singularity, so its worldvolume is a big-bang/big-crunch cosmology. In the Lorentzian spacetime the brane is causally separated from the asymptotic boundary of the spacetime. The usual holographic dictionary implies that the bulk spacetime is entirely encoded in the dual field theory, but the braneworld is thus encoded in a complicated, non-local way.

The nature of this encoding can be understood by analytic continuation to Euclidean signature. In the Euclidean solution the brane intersects the boundary. Following \cite{Takayanagi:2011zk,Fujita:2011fp}, this intersection can be understood as a holographic description of a BCFT. The state dual to the bulk black hole spacetime is then obtained by the CFT path integral with the corresponding BCFT boundary condition. 

\begin{figure}
    \centering
    \includegraphics[width=1.0\linewidth]{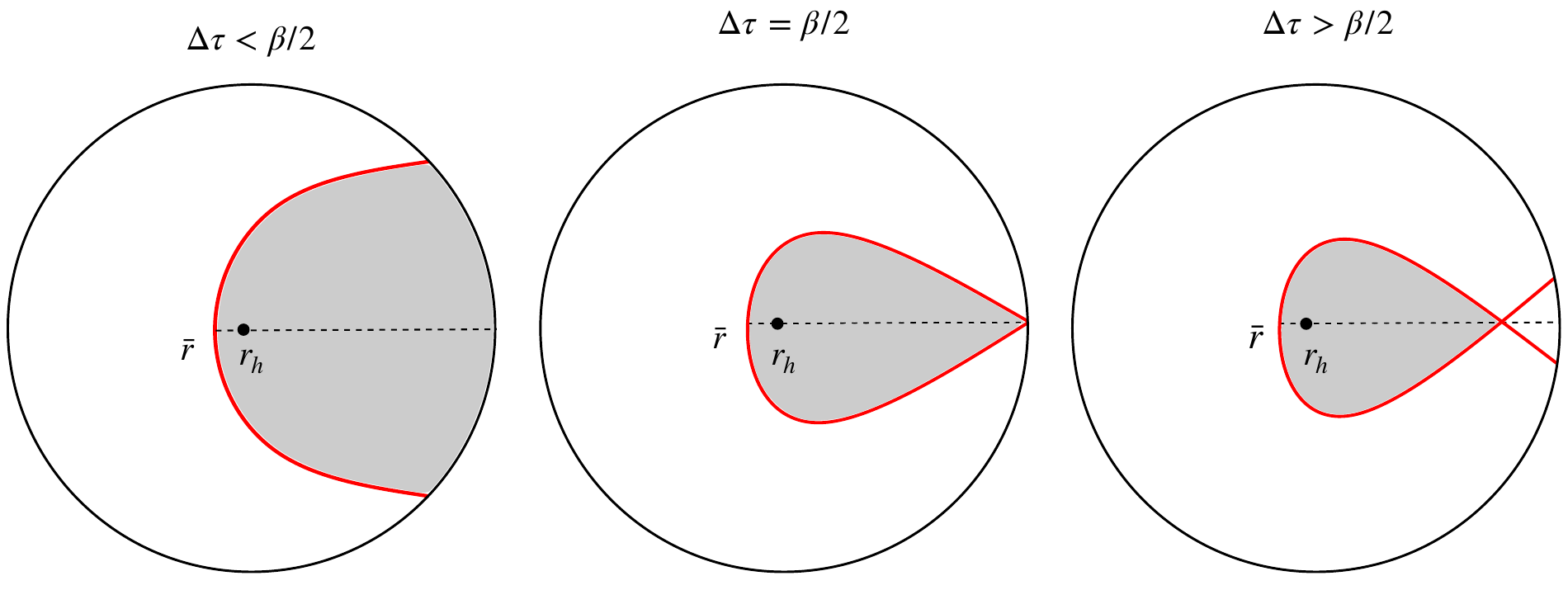}
    \caption{Euclidean ETW brane trajectories for different intersection points at the boundary. The shaded regions are the only bulk parts we keep from the original Euclidean geometry without the brane. The black dashed line is the $t=0$ slice. For $\Delta \tau <\beta/2$, the brane hits the asymptotic boundary at times $t>0$; for $\Delta \tau=\beta/2$, the brane meets at the asymptotic boundary at $t=0$. In these cases, we can construct the dual CFT state of the $t=0$ slice through the Euclidean path integral. However, for $\Delta \tau>\beta/2$, the brane self-intersects before it reaches the asymptotic boundary, implying our inability to construct the CFT state dual to the $t=0$ slice through the Euclidean path integral.}
    \label{EuclidConst}
\end{figure}

However, in simple examples with pure tension branes, this Euclidean construction fails in a certain regime. To model realistic cosmologies, we would like the dynamics to be described by an approximately local theory on the brane worldvolume. Such an approximately local description is obtained when we tune the brane tension $T \approx \frac{1}{L}$, where $L$ is the bulk AdS scale. This makes the induced cosmological constant on the brane small, so the associated AdS scale $\ell \gg L$ \cite{Randall:1999ee,Randall:1999vf,Karch:2000ct}. We refer the reader to \cite{Bueno:2022log,Panella:2024sor} for a detailed derivation of the worldvolume effective action. The theory on the brane is then approximately local on scales large compared to $L$ (we take $L \gg l_{pl}$ so the bulk theory remains classical). But as we tune the brane tension, for $d>2$ the time $\Delta \tau$ at which the brane hits the boundary increases, and for a given bulk black hole solution it will eventually exceed $\beta/2$, where $\beta$ is the period in the Euclidean black hole, see figure \ref{EuclidConst}. Increasing the brane tension, the brane will self-intersect before it reaches the Euclidean boundary, and we do not have a controlled Euclidean construction giving the state of interest. 

There are setups where this problem can be avoided: \cite{Antonini:2019qkt} showed that it could be avoided by considering a charged black hole and tuning the charge towards extremality as we tune the tension,\footnote{See also \cite{Antonini:2024bbm} for a recent discussion of a magnetically charged case.} and \cite{Ross:2022pde}  showed that it could be avoided by considering a brane with hyperbolic spatial slices. However, these can seem like rather special models, and each have their drawbacks. In the former case the bulk solution is fine tuned, corresponding to some rather special matter content on the brane in the dual description. In the latter case, the bulk solution is expected to be meta-stable to quantum tunneling producing branes in the region outside the black hole which escape to the conformal boundary in the Lorentzian solution.

In this work, we consider generalising the theory by including additional matter on the braneworld worldvolume, not just considering a pure tension brane. We focus on the regime where $T \approx \frac{1}{L}$, where we have an approximately local theory on the brane. We consider first a perfect fluid stress tensor on the brane. To have a turning point, we need to have sufficient matter density at the turning point. Assuming there is a turning point, we find that for any matter content with equation of state parameter $w > -1$, $\Delta \tau \propto \frac{\ell L}{\bar r}$, where the constant of proportionality is of order 1, and $\bar r$ is the minimal radius of the brane in the Euclidean solution. With no spatial curvature, $\beta = \frac{4\pi L^2}{d r_h}$, where $r_h$ is the horizon radius, so we can avoid self-intersection if $\bar r/\ell > C r_h/L$, where $C$ is some order one factor. We thus need sufficient density of matter on the brane and a large enough turning point, but this does not require fine tuning or a special choice of matter. In this context it is thus the self-intersection problem that seems special, appearing only when we consider pure tension branes.  We also consider turning on classical fields on the brane. We find that turning on an axion can resolve the self-intersection problem (indeed, a homogeneous axion is just stiff matter with $w=1$, so it is included in the preceding perfect fluid analysis), while a massive scalar field doesn't help. 

In \cite{Ross:2022pde}, it was observed that the Euclidean worldvolume of the brane is a spacetime wormhole connecting two asymptotically AdS ends, and in the limit where the effective theory becomes approximately local on the brane, the Euclidean construction here is related to the construction of Euclidean wormhole solutions in gravity. Indeed, in the case where the spatial slices are hyperbolic, as we tune the brane effective cosmological constant to zero the brane worldvolume geometry approaches the Maldacena-Maoz wormhole solution \cite{Maldacena:2004rf}. The observation that the self-intersection problem can be avoided by adding matter is consistent with recent constructions of wormholes supported by heavy particles \cite{Chandra:2022bqq}. (The observation that wormholes can be supported by a perfect fluid stress tensor was also made in \cite{Maloney:2025tnn}, which appeared while this paper was in preparation.) This viewpoint also illuminates why an axion is helpful but a scalar field is not.

In the next section, we consider the models with perfect fluid matter on the brane. In section \ref{scalar}, we consider a massive scalar field on the brane, and see that this does not resolve the self-intersection problem. Then in section \ref{axion}, we consider an axion field on the brane, and see that this can also resolve the self-intersection problem. We conclude with some discussion in section \ref{disc}. 

\section{Perfect fluid matter}
\label{fluid}

We consider a model in $d+1$ dimensions with coordinates $x^\mu$, with an Einstein-Hilbert action in the bulk with cosmological constant $\Lambda = - \frac{d(d-1)}{2 L^2}$, and a $d$-dimensional brane with tension $T$ and a perfect fluid stress tensor on the brane. We consider a vacuum black hole solution in the bulk, \begin{equation}\label{ambient}
ds^2 = -f(r) \,dt^2 + {dr^2 \over f(r)} + r^2\,d\Sigma^2\;, \qquad f(r) = k + \frac{r^2}{L^2} - {\mu \over r^{d-2}}\; , 
\end{equation}
where the spatial sections $d\Sigma_k^2$ have curvature $k=\pm 1, 0$. The Euclidean solution obtained by analytically continuing $t \to - i \tau$ has periodicity
\begin{equation}
\beta = \frac{4\pi r_h L^2}{d r_h^2 + (d-2) k L^2}, 
\end{equation}
 where $f(r_h)=0$. We take the brane to move along some trajectory ($t(\lambda), r(\lambda)$). The $d$-dimensional coordinates along the brane are $x^a$. The dynamics of the brane in this background is determined by the Israel junction conditions, which relate the extrinsic curvature $K_{ab}$ to the brane stress-energy tensor, 
\begin{equation}\label{junction}
K^a\;_b - K\;\; \delta^a\;_b = 8\pi G\,S^a\;_b- (d-1) T \delta^a_{\ b} \;, \qquad S^a\;_b= (p+\rho)\,U^a U_b + p\; \delta^a\;_b\;, 
\end{equation}
where $U=\partial_\lambda$ is the four-velocity of the fluid in terms of the shell's proper time $\lambda$.

The induced metric on the brane is 
\begin{equation} \label{induced}
ds^2_{\rm EOW} = -\left(f {\dot t}^2 - {{\dot r}^2 \over f}\right) d\lambda^2 + r^2 d\Sigma_k^2 =  - d\lambda^2 + r^2 d\Sigma_k^2\; ,
\end{equation}
where the dot indicates $d/d\lambda$, and in the second step we choose the parameter to be proper time along the brane, implying that $f \dot t^2 - \frac{\dot r^2}{f} = 1$. The explicit form of the non-trivial components of the extrinsic curvature is then 
\begin{equation}\label{extrinsic}
K^\lambda\;_\lambda = {\dot H \over {\dot r}} \;, \qquad K^\Sigma\;_\Sigma = {H \over r}\;,  \qquad K = K^a\;_a = {{\dot H} \over {\dot r}} + (d-1) {H \over r},
\end{equation}
where 
\begin{equation}\label{H}
H = \sqrt{f(r) + {\dot r}^2}.
\end{equation}
The non-trivial dynamical equations are the timelike and angular components
\begin{equation}
K^\lambda\;_\lambda - K = -8\pi G\, \rho - (d-1) T\;, \qquad K^\Sigma\;_\Sigma - K = 8\pi G\; p - (d-1) T\;.
\end{equation}
The timelike equation is of first order
\begin{equation}\label{jun}
H= \sqrt{f+ {\dot r}^2} = \left( T + {8\pi G \over d-1} \,\rho\right) \,r\;,
\end{equation}
whereas the angular equation is of second order. However, the latter equation is always satisfied provided the timelike equation is satisfied and the dynamics of the perfect fluid is adiabatic, i.e.
\begin{equation}\label{adiabatic}
p {d\over d \lambda} \,r^{d-1} = - {d \over d\lambda} \left( \rho \,r^{d-1}\right)\;.
\end{equation}
Therefore, it suffices to deal with the first-order timelike equation, which we can write as
\begin{equation}\label{embeding}
{\dot r}^2 = H^2 - f =  \left( T+ {8\pi G \over d-1} \rho\right)^2 r^2 - f(r).
\end{equation}
We note that this only depends on the energy density. 

We are particularly interested in the limit where the dynamics is well-approximated by a local theory on the brane. This arises when $T \approx \frac{1}{L}$, so that the induced cosmological constant on the brane $\Lambda_{brane} = - \frac{(d-1)(d-2)}{\ell^2}$, where $\frac{1}{\ell^2} = \frac{1}{L^2} - T^2$, has $\ell  \gg L$, and all other relevant length scales on the brane are large compared to $L$. This can be seen at the level of the dynamical equation \eqref{embeding}, which becomes
\begin{equation} \label{Friedmann}
\frac{{\dot r}^2}{r^2} = \frac{16\pi G_{brane}TL}{(d-2)(d-1)} \rho  - \frac{k}{r^2} - \frac{1}{\ell^2}  + \frac{\mu}{r^d} + \left( \frac{8\pi G_{brane} L \rho }{(d-2)(d-1)}\right)^2,
\end{equation}
Where $G_{brane} = (d-2) G/L$ is the Newton constant in the effective theory on the brane. In a regime where $G_{brane} \rho \ll \frac{1}{L^2}$, the last term is negligible compared to the first, and this reduces to the usual Friedmann equation on the brane (with the mass of the bulk black hole interpreted as conformal matter on the brane).

Using the condition for $\lambda$ to be proper time to determine $\dot t$ in terms of $\dot r$, we can rewrite \eqref{embeding} as 
\begin{equation}\label{param}
\left({dt \over dr}\right)^2 = \left({{\dot t} \over {\dot r}}\right)^2  = \frac{1}{f^2} + \frac{1}{f \dot r^2} = {H^2 \over f^2 (H^2-f)}
\;.
\end{equation}
In the Euclidean solution this becomes 
\begin{equation}
\left({d\tau \over dr}\right)^2 = {H^2 \over f^2 (f-H^2)}
\;.
\end{equation}
In the Euclidean solution the brane has a minimum radius at the turning point where $r = \bar r$. At the turning point $\dot r=0$, so  $f({\bar r} ) = H({\bar r})^2$. We take this to be at $\tau = \beta/2$, as the brane is inside the black hole horizon, and we take $\tau = 0$ to correspond to the asymptotic region outside the horizon. The brane then reaches the boundary at
\begin{equation}\label{bound}
\tau = {\beta \over 2} \pm \int_{\bar r}^\infty {dr \over f} {H \over \sqrt{f-H^2}}\; , \qquad H = \left( T + {8\pi G \over d-1} \,\rho\right) r\;.
\end{equation}
This equation is identical to the pure-tension equation, but we have shown it is valid for any type of perfect fluid with adiabatic evolution on the world-volume. To avoid self-intersection we want
\begin{equation}\label{delta}
\Delta \tau = \int_{\bar r}^\infty {dr \over f} {H \over \sqrt{f-H^2}}< \frac{\beta}{2} \; .
\end{equation} 

We are interested in studying the regime of small $L$, where we have an approximately local theory on the brane. This is where the self-intersection problem arises in the simple pure tension case. In this regime we can take $H \approx Tr$ and $f \approx r^2/L^2$, while $\sqrt{f-H^2} = \dot r$ is given by the Euclidean version of \eqref{Friedmann}, 
\begin{equation} \label{Fr2}
\frac{{\dot r}^2}{r^2} \approx \frac{k}{r^2} + \frac{1}{\ell^2} - \frac{16\pi G_{brane}}{(d-2)(d-1)} \rho  -   \frac{\mu}{r^d} . 
\end{equation}
Thus
\begin{equation}
    \Delta \tau \approx L \int_{\bar r}^\infty \frac{dr}{r \dot r}. 
\end{equation}
The remaining integral does not scale with $L$, so if we want this to be smaller than $\beta$, we also need to take $r_h \sim L$. Then $\mu \sim L^{d-2}$, and the $\mu$ term in the Friedmann equation \eqref{Fr2} is negligible. We see that we then need other forms of matter on the brane (or $k=-1$) to have a non-trivial turning point in this regime. If we consider for simplicity a brane with $k=0$, $\dot r(\bar r)=0$ determines the density at the turning point, $\bar \rho = \rho(\bar r)$  as
\begin{equation}
  \frac{16\pi G_{brane}}{(d-2)(d-1)} \bar \rho =  \frac{1}{\ell^2}. 
\end{equation}
Setting $x = r/\bar r$, we then have
\begin{equation} \label{rhogen}
    \Delta \tau = \frac{\ell L}{\bar r} \int_1^\infty \frac{dx}{x^2 \sqrt{1- \rho(r)/\bar \rho}}. 
\end{equation}
We can achieve $\Delta \tau < \frac{\beta}{2}$ for $r_h \sim L$ by taking $\bar r/\ell$ larger than some order one value.

For non-zero $k$, we have instead 
\begin{equation}
    \Delta \tau = \frac{\ell L}{\bar r} \int_1^\infty \frac{dx}{x^2 \sqrt{1 + k\ell^2/(\bar r^2 x^2) - (1 + k \ell^2/\bar r^2) \rho(r)/\bar \rho}}.
\end{equation} 
This makes a quantitative difference to the relation between $\Delta \tau$ and $\bar r/\ell$ but doesn't affect the qualitative conclusion.

A nice example to consider is dust, as this has a natural interpretation from the dual field theory perspective. We can add a shell of dust on the boundary by acting with a collection of local operator insertions for some heavy field with conformal dimension $\Delta \gg 1$, as in \cite{Chandra:2022bqq,Sasieta:2022ksu}. Dust has $\rho \sim r^{-(d-1)}$, so 
\begin{equation}
    \Delta \tau = \frac{\ell L}{\bar r} \int_1^\infty \frac{dx}{x^2 \sqrt{1- x^{-(d-1)}}} = \frac{\ell L}{\bar r} \sqrt{\pi} \frac{ \Gamma(\frac{d}{d-1})}{\Gamma(\frac{d+1}{2(d-1)})}.
\end{equation}
We could also consider other forms of matter on the brane, and more complicated models with multiple components. The key point is that the addition of brane matter enables us to solve the turning point equation in the small $L$ regime, and to choose a large value of $\bar r$, without changing $r_h$.\footnote{A similar separation was achieved by a different mechanism in the charged black hole proposal of \cite{Antonini:2019qkt}.}

The addition of matter on the brane provides an attractive resolution of the self-intersection problem: we just need to have a sufficient amount of matter, the value is not fine-tuned, and as the bulk black hole solution is unmodified we would not expect this to introduce any new instabilities in the dual theory, unlike the curvature case studied in \cite{Ross:2022pde}. The addition of matter can be given a microscopic description, as discussed above for the case with dust. However, this involves turning on a relatively complicated source on the boundary, with a large number of local operator insertions. It's therefore interesting to ask whether we can also construct braneworlds that avoid the self-intersection problem by turning on classical fields on the brane, corresponding to simpler boundary sources. In the next section we see that turning on a massive scalar field doesn't work, and then in section \ref{axion} we will see that turning on an axion does work. This is also consistent with expectations from spacetime wormhole constructions. 

\section{Scalar field}
\label{scalar}

In cosmology one often considers coupling to a scalar field $\phi$ with potential $V(\phi)$. Braneworld scenarios with a scalar field on the brane were considered in \cite{Kanda:2023zse,Fujiki:2025yyf}. If we consider a spatially homogeneous but time-varying scalar $\phi(\lambda)$, this can be thought of as a particular case of the preceding perfect fluid analysis, with 
\begin{equation}\label{scalarf}
p_\phi = {1\over 2} {\dot \phi}^{\,2} - V(\phi) \;, \qquad \rho_\phi = {1\over 2} {\dot \phi}^{\,2} + V(\phi)\;,
\end{equation} 
where ${\dot \phi} \equiv d\phi /d\lambda$. However, a scalar field on its own cannot resolve the self-intersection problem. We will first show this in general, without restricting to the regime of small $L$. 

The dynamical equations in Lorentzian signature are 
\begin{equation}\label{embed}
H \equiv \sqrt{f(r) + {\dot r}^{\,2}} = \left( T + {8\pi G\over d-1} \,\rho_\phi \right) r  =  T r+ {8\pi G \over d-1} \,r\, \left({{\dot \phi}^{\,2} \over 2} + V(\phi) \right) \;, \qquad {\dot t}^{\,2} = {H^2 \over f^2} \;, 
\end{equation} 
and the worldvolume scalar equation
\begin{equation} \label{scL}
{\ddot \phi} + {d-1 \over r} \,{\dot r} \, {\dot \phi} = - \partial_\phi V(\phi)\;, 
\end{equation} 
which is equivalent to the adiabatic condition and thus ensures that equations (\ref{embed}) and (\ref{scL}) are complete. 
To see this, we can just plug the forms (\ref{scalarf}) into the adiabatic condition (\ref{adiabatic}), and check that it is satisfied provided (\ref{scL}) is satisfied. 
\begin{figure}
    \centering
    \includegraphics[width=0.75\linewidth]{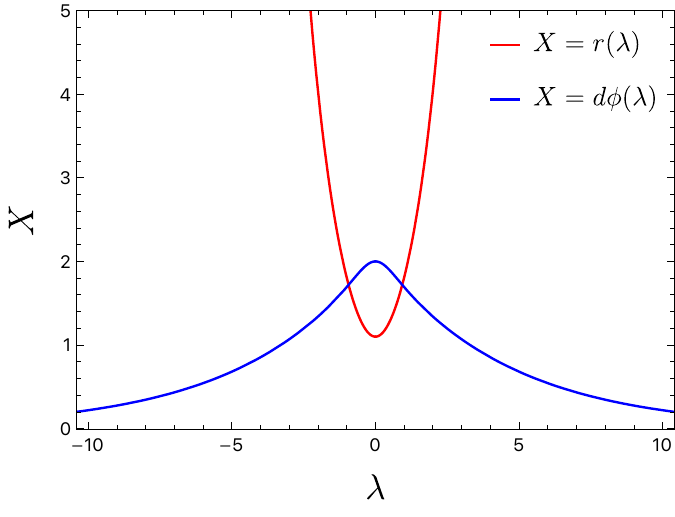}
    \caption{Scalar and the brane profile for the potential $V(\phi)=m^2\phi^2$ with parameters $m^2=-1/4$, $d=4$, $\phi(\lambda_t)=1/2$, $\bar{r}=11/10$ and $r_h=1$. $\lambda_t$ is the turning point for the brane and the scalar field. For the clarity of the plot we multiply $\phi(\lambda)$ with the spatial dimension $d$.}
    \label{Euclidsca}
\end{figure}
Passing to the Euclidean signature $\lambda \rightarrow -i\lambda$, and $t\rightarrow -i\tau$ we get the analogous Euclidean equations 
\begin{equation}\label{embede}
H \equiv \sqrt{f(r) - {\dot r}^{\,2}} = \left( T+ {8\pi G\over d-1} \,\rho_\phi \right) r = T r  + {8\pi G \over d-1} \, \left(-{{\dot \phi}^{\,2} \over 2} + V(\phi) \right) \, r\;, \qquad {\dot \tau}^{\,2} = {H^2 \over f^2} \;, 
\end{equation} 
with the Euclidean scalar equation obtained by the usual sign-flip of the potential with respect to the Lorentzian one: 
\begin{equation} \label{scE}
{\ddot \phi} + {d-1 \over r} \,{\dot r} \, {\dot \phi} =  \partial_\phi V(\phi)\;.
\end{equation} 
The term linear in ${\dot \phi}$ is a `cosmological' friction term with effective Hubble constant ${\dot r}/r$. To have a time-symmetric Euclidean solution which we can analytically continue to obtain initial data for the Lorentzian evolution, we must have $\dot \phi=0$ at the turning point $r=\bar r$ where $\dot r=0$, see figure \ref{Euclidsca}.

\begin{figure}
    \centering
    \includegraphics[width=1\linewidth]{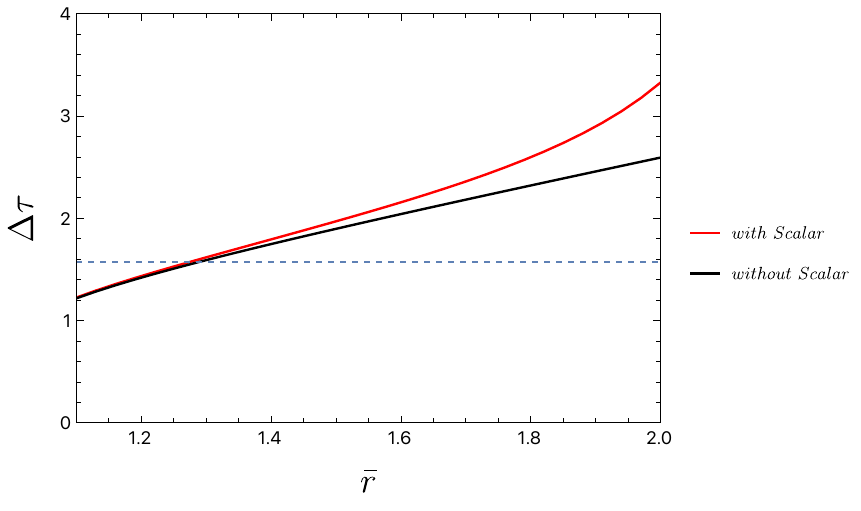}
    \caption{$\Delta \tau$ vs brane turning point $\bar{r}$ for the scalar potential $V(\phi)=m^2\phi^2$ with parameters $m^2=-1/4$, $d=4$, $\phi(\lambda_t)=1/2$, and $r_h=1$. The dashed line is $\Delta \tau = \beta/2$. The value of $\Delta \tau$ with the scalar field is lower bounded by the pure tension value. The Euclidean ETW brane self-intersects when $\Delta \tau$ crosses $\beta/2$.}
    \label{fig:scalar}
\end{figure}

In the absence of the friction term, the Euclidean energy $\rho_\phi =-{{\dot \phi}^{\,2} \over 2} + V(\phi)$ would be conserved. The cosmological friction implies
\begin{equation}
    \dot \rho_\phi = (- \ddot \phi + \partial_\phi V(\phi)) \dot \phi = \frac{d-1}{r} \dot r \dot \phi^2 \geq 0, 
\end{equation}
so along the evolution of the brane $\rho_\phi (r) \geq {\bar \rho 
 }$, which implies 
\begin{equation}
 H(r) = \left( T + {8\pi G \over d-1} \rho_\phi (r) \right) r \geq {\bar H} \equiv  \left( T + {8\pi G \over d-1} \rho_\phi (\bar r) \right) \bar r \;.   
\end{equation}
This means that the elapsed time along the brane will satisfy the inequality
\begin{equation}
    \Delta \tau = \int_{\bar r}^\infty {dr \over f(r)} {H(r) \over \sqrt{f(r) - H(r)^2}} \geq \int_{\bar r}^\infty {dr \over f(r)} {{\bar H} \over \sqrt{f(r) - {\bar H}^2}} = {\overline{\Delta \tau}}\;,
\end{equation}
where ${\overline{\Delta \tau}}$ is the time for a pure-tension brane with the same turning point ${\bar r}$. That is, adding a scalar field on the brane can only {\it increase} the amount of time it takes to reach the boundary. Thus, since a pure-tension brane has a self-intersection problem above a critical value of $\bar r$, so will the brane with a scalar field. This is confirmed by numerical analysis; see the plot in figure \ref{fig:scalar}. 

This gives a simple general argument that a scalar on its own is insufficient. One might however wonder how the previous analysis in the small $L$ limit for a general perfect fluid fails in this case. The problem is simply that we can't build a wormhole using just a scalar field. In the Euclidean analysis, we saw that $\frac{d \rho_\phi}{dr} >0$, which implies that $\rho_\phi(r) > \bar \rho$ for $r > \bar r$, so the square root in \eqref{rhogen} has a negative argument. What's happening is that we have 
\begin{equation}
    \dot r^2 \approx k +  \frac{r^2}{\ell^2} - \frac{16\pi G_{brane}}{(d-1)(d-2)} \rho_\phi r^2,
\end{equation}
which implies 
\begin{equation}
    2\dot r \ddot r \approx  \left(\frac{1}{\ell^2}- \frac{16\pi G_{brane}}{(d-1)(d-2)} \rho_\phi   \right) 2 r \dot r - \frac{16\pi G_{brane}}{(d-1)(d-2)} \dot \rho_\phi r^2 , 
\end{equation}
and $\dot \rho_\phi = \frac{d-1}{r} \dot r \dot \phi^2$, so at the turning point  
\begin{equation}
    \ddot r \approx - \frac{k}{r} - \frac{8\pi G_{brane}}{(d-2)} \dot \phi^2 r.
\end{equation}
Thus, unless $k=-1$, we have $\ddot r<0$, and the brane isn't moving out towards the boundary at the turning point. 

\section{Axion field}
\label{axion}

We now show that adding an axion field on the brane could resolve the self-intersection problem. The use of axions as a matter source in the context of spacetime wormholes has a long history, going back to the original work of Giddings and Strominger \cite{Giddings:1987cg,Giddings:1988cx}. Axion wormholes can also be constructed in AdS/CFT, see e.g.  \cite{Arkani-Hamed:2007cpn,Hertog:2017owm}. Axions are massless pseudoscalar fields, dual in $d$ spacetime dimensions to $d-2$ form potentials. In the analytic continuation to Euclidean signature, if we keep the usual sign for the kinetic term of the $d-1$ form field strength, the sign of the axion kinetic term is reversed, so a spatially homogeneous axion $\chi(\lambda)$ corresponds to a perfect fluid with the Euclidean pressure and energy density
\begin{equation}\label{scalarf}
p_\chi = {1\over 2} {\dot \chi}^{\,2}  \;, \qquad \rho_\chi = {1\over 2} {\dot \chi}^{\,2} \;,
\end{equation} 
and the equation of motion 
\begin{equation} \label{scL}
{\ddot \chi} + {d-1 \over r} \,{\dot r} \, {\dot \chi} = 0. 
\end{equation} 
This thus corresponds to stiff matter, with $w=1$. Indeed, we can integrate the axion equation of motion to obtain $\dot \chi = \frac{c}{r^{d-1}}$, so $\rho_\chi = \frac{c^2}{2 r^{2(d-1)}}$. Because the axion is a pseudoscalar, the condition for a time-symmetric Euclidean solution is that $\chi=0$ at the turning point $r = \bar r$, with $\dot{\bar 
\chi} = \dot \chi(\bar r)= \frac{c}{\bar r^{d-1}}$ as free data.

\begin{figure}
    \centering
    \includegraphics[width=0.75\linewidth]{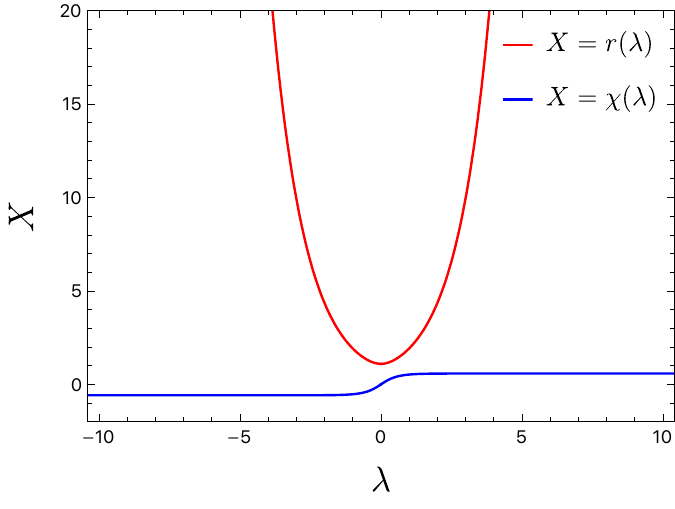}
    \caption{Axion and brane profile for $\dot \chi(\lambda_t)=1$, $\chi(\lambda_t)=0$, $\bar{r}=11/10$ and $r_h=1$. $\lambda_t$ is the turning point for the brane. }
    \label{fig:axionprof}
\end{figure}

If we consider a model with just an axion, we have a similar analysis to the dust model in section \ref{fluid}. For $k=0$, the turning point condition fixes $\bar \rho_\chi = \frac{1}{2} \dot{\bar \chi}^2$ by 
\begin{equation}
  \frac{16\pi G_{brane}}{(d-2)(d-1)} \bar \rho_\chi =  \frac{1}{\ell^2},
\end{equation}
and setting $x = r/\bar r$, we then have
\begin{equation}
    \Delta \tau = \frac{\ell L}{\bar r} \int_1^\infty \frac{dx}{x^2 \sqrt{1- x^{-2(d-1)}}} .
\end{equation}
We can have $\Delta \tau < \beta/2$ if $\bar r/\ell$ is sufficiently large. 

It is also interesting to compute the difference in value of the axion between the two ends of the brane. The profile for the axion field along the brane is shown in figure \ref{fig:axionprof}. We see that the axion approaches a constant value in the asymptotic region, as expected for a massless (pseudo)scalar in an asymptotically AdS background. The axion theory has a shift symmetry, so the value of this constant field is not really physically significant; the significant quantity is the difference between the value on the two ends. This acts as a source for the corresponding operator in the dual theory. We have 
\begin{equation} \label{dchi}
    \Delta \chi = 2 \int_{\bar r}^\infty \frac{\dot \chi}{\dot r} dr = 2\dot{\bar \chi} \ell \int_1^\infty \frac{dx}{x^d \sqrt{1 - x^{-2(d-1)}}} = \sqrt{\frac{\pi (d-2)}{8 (d-1) G_{brane}}} .
\end{equation}
We see that we get a constant value of the difference, independent of the choice of bulk solution, as noted previously in discussions of axion wormholes, e.g. \cite{Arkani-Hamed:2007cpn}.\footnote{The analysis in \cite{Arkani-Hamed:2007cpn} has $k=1$, and works in flat space, so the numerical factor is different -- but by coincidence, not very different: they have instead $\Delta \chi =\sqrt{\frac{\pi (d-1)}{8 (d-2) G_{brane}}}  $.}
This implies that if we have a pure axion configuration there is an unexpected mismatch between the bulk parameters and the boundary conditions; we have a free bulk parameter $\dot{\bar \chi}$ for the axion but not a corresponding free parameter in the boundary data.  This condition on $\Delta \chi$ is  relaxed if we consider a model with an axion and other matter. Consider for example a model with an axion and dust. Then $\rho = \rho_d + \rho_\chi$, and we have
\begin{equation}
    \Delta \chi = \frac{\dot{\bar \chi}}{\sqrt{\bar \rho_d+ \bar \rho_\chi}} \sqrt{\frac{(d-2)(d-1)}{4 \pi G_{brane}}} \int_1^\infty \frac{dx}{x^d \sqrt{1 - \rho(r)/\bar\rho}}.
\end{equation}
with
\begin{equation}
    \frac{\rho}{\bar \rho} = \frac{\bar \rho_d}{\bar \rho_d + \bar \rho_\chi} x^{-(d-1)}+ \frac{\bar \rho_\chi}{\bar \rho_d + \bar \rho_\chi} x^{-2(d-1)}.
\end{equation}
Thus, if $\bar \rho_\chi$ is a small contribution to the density at the turning point, we have $\Delta \chi \sim \dot{\bar \chi}$. We can thus have small values of $\Delta \chi$ if the brane configuration is supported by other matter. As we increase $\dot{\bar \chi}$, $\Delta \chi$ saturates at the previous value. We could also have configurations where $\Delta \chi$ exceeds that value by including matter that contributes negatively to $\rho$, such as a massless scalar. Thus in general considering more complicated theories restores the usual match between the set of free parameters in the bulk and the boundary conditions. We can in general choose a value of $\Delta \chi$, which determines $\dot{\bar \chi}$. If we tune $\Delta \chi$ to the critical value \eqref{dchi}, the axion becomes dominant over other forms of matter in the bulk evolution, that is $\dot{\bar \chi} \to \infty$.

\section{Discussion}
\label{disc}

In this paper, we extended earlier discussions of braneworld cosmology inside black holes by considering additional matter on the brane. This is a natural extension to consider in general for modelling more realistic cosmologies. We considered it primarily as a solution to the self-intersection problem seen in \cite{Cooper:2018cmb} when the braneworld cosmology model was extended to approximately local theories on the brane in $d>2$. We saw that for  perfect fluid matter with sufficient energy density $\bar \rho$ on the brane and large enough $\bar r$, the self-intersection problem is resolved. To have $\bar r \sim \ell$, we need to choose $r_h \sim L$, so the bulk black hole is of order the bulk AdS scale, and does not make a significant contribution to the matter on the brane. This resolution has the advantage over previous mechanisms \cite{Antonini:2019qkt,Ross:2022pde} that it does not require fine tuning and it does not modify the bulk solution. 

We thus see that it's actually fairly easy to resolve this self-intersection problem. The Euclidean brane is an AdS spacetime wormhole connecting two asymptotic boundaries, so this also shows that we can construct such spacetime wormhole solutions supported by perfect fluid matter. This was also noted in the recent work \cite{Maloney:2025tnn}, which appeared while our paper was in preparation. 

The resolution is however not universal. We considered a homogeneous scalar field on the brane, which was also considered recently in \cite{Kanda:2023zse,Fujiki:2025yyf}, and found that this on its own would not give a suitable brane solution: in the small $L$ regime we can't build a wormhole using just a homogeneous scalar field.\footnote{In \cite{Maloney:2025tnn}, inhomogeneous scalar fields were considered. It would be interesting to also extend our discussion to this case.} By contrast, an axion field on the brane can resolve the self-intersection problem. If we consider an axion field on its own, there is a mismatch between the set of free data in the bulk and the set of free parameters in the boundary data, but this mismatch is resolved by considering more general theories with additional matter on the brane in addition to the axion. 

An important direction for future work is to understand the CFT states dual to these bulk geometries, and the encoding of the cosmology in them. For some choices of matter, such as dust, there is a simple microscopic description of the boundary conditions. Even in the pure tension case there has been relatively little investigation of the corresponding states obtained by the Euclidean path integral in the CFT and their relation to the cosmology, and this will clearly be more complicated in the present case with additional matter on the brane. However states constructed by shells of dust have recently been studied in the context of black hole microstates \cite{Balasubramanian:2022gmo,Balasubramanian:2022lnw,Geng:2024jmm}, and there some universal behaviours were found in a limit where the shell stays near the boundary, which is related to the small $L$ limit considered here, so it may be that further progress can be made along those lines in this regime. 

\section*{Acknowledgements} 
We thank Jan de Boer, Martin Sasieta and Eduardo Velasco-Aja for useful discussions. The work of SFR is supported in part by STFC through grant number ST/X000591/1. JLFB, AKP and JFP are supported by the Spanish Research Agency via grants CEX2020-001007-S and PID2021-123017NB-I00, funded by MCIN/AEI/10.13039/501100011033, by the ‘Atracci\'on de Talento’ program (Comunidad de Madrid) grant 2020-T1/TIC-20495, and by ERDF `A way of making Europe.' JLFB wishes to thank KITP at Santa Barbara for hospitality during the completion of this work,  supported in part by grant NSF PHY-2309135 to the Kavli Institute for Theoretical Physics (KITP). AKP further acknowledges support from COST Action CA22113 via an STSM grant, which funded a visit to Durham University during which part of this work was carried out.

\bibliographystyle{JHEP}
\bibliography{draft}

\end{document}